**Temperature programed desorption of water ice from the surface of amorphous carbon and silicate grains as related to planet-forming disks**

Alexey Potapov[1], Cornelia Jäger[1], and Thomas Henning[2]

[1]*Laboratory Astrophysics Group of the Max Planck Institute for Astronomy at the Friedrich Schiller University Jena, Institute of Solid State Physics, Helmholtzweg 3, 07743 Jena, Germany, email: alexey.potapov@uni-jena.de*

[2]*Max Planck Institute for Astronomy, Königstuhl 17, D-69117 Heidelberg, Germany*

**Abstract**

Understanding the history and evolution of small bodies, such as dust grains and comets, in planet-forming disks is very important to reveal the architectural laws responsible for the creation of planetary systems. These small bodies in cold regions of the disks are typically considered as mixtures of dust particles with molecular ices, where ices cover the surface of a dust core or are actually physically mixed with dust. Whilst the first case, *ice-on-dust*, has been intensively studied in the laboratory in recent decades, the second case, *ice-mixed-with-dust*, present uncharted territory. This work is the first laboratory study of the temperature-programmed desorption (TPD) of water ice *mixed with* amorphous carbon and silicate grains. We show that the kinetics of desorption of $H_2O$ ice depends strongly on the dust/ice mass ratio, probably, due to the desorption of water molecules from a large surface of fractal clusters composed of carbon or silicate grains. In addition, it is shown that water ice molecules are differently bound to silicate grains in contrast to carbon. The results provide a link between the structure and morphology of small cosmic bodies and the kinetics of desorption of water ice included in them.

1. **Introduction**

Interstellar and circumstellar ices are considered to be possible sources of organic molecules responsible for the origin and early evolution of life on Earth (Oro 1961; Cronin & Chang 1993; Brack 1999; Pearce et al. 2017). A number of amino acids have been found in comets and meteorites (Cronin & Chang 1993; Elsila, Glavin, & Dworkin 2009; Altwegg et al. 2016) and have been produced in interstellar ice analogues in the laboratory (Bernstein et al. 2002; Muñoz Caro et al. 2002; Nuevo et al. 2006; Elsila et al. 2007). On this background, there is an increasing interest to study the chemistry, the structure, and the dynamical properties of laboratory cosmic ice analogues in the last decades.

In the interstellar medium (ISM), ices form a mantle around a dust core (Allamandola et al. 1999) through condensation of gaseous species onto cold dust grains. Dust grains (ice-covered or not), which are mainly carbon or silicate based particles (Henning & Salama 1998; Draine 2003), represent the most pristine starting material for planetary systems. Ice plays an important role in the grain growth. According to coagulation models (Wada et al. 2009; Wettlaufer 2010) and collisional experiments (Gundlach & Blum 2015), ice-coated grains are expected to stick together much more efficiently. After a coagulation of dust grain monomers covered by ice in dense regions, such as dense molecular clouds and planet-forming disks, ice can be trapped inside grain aggregates. Laboratory experiments on dust particle aggregation by collisions [see, for example, (Wurm & Blum 1998; Krause & Blum 2004) and a recent review (Blum 2018)] and an analysis of the fractal dust particles observed by Rosetta (Fulle & Blum 2017) have shown that final dust aggregates are highly porous. Ice can cover the surface of grains and grain aggregates and can fill the pores of the aggregates via adsorption of volatile molecules from the gas phase. As it was shown by the examination of the samples returned by the Stardust mission, there are indications that the dust and water-ice agglomerates were mixed before cometesimals formed in the outer solar system (Brownlee et al. 2006). The results of the Rosetta mission demonstrated that the nucleus of the comet 67P/Churyumov-Gerasimenko is characterized by an average dust-to-ices mass ratio of 7.5 consistent with a mixture of ices, Fe-sulphides, silicates, and hydrocarbons (Fulle et al. 2017). It was also recently shown in laboratory experiments that a part of water ice molecules mixed with silicate grains at 8 K is trapped on silicate agglomerates and does not desorb up to 200 K (Potapov et al. 2018). This trapped ice can survive the transfer from a dense molecular cloud into a protoplanetary disk, even at high temperatures. Thus, it is very probable that dust grains and comets in planet-forming disks present, as a rule, physical mixtures of ice and dust.

Observations and dedicated laboratory experiments have shown that the main constituent of interstellar and circumstellar ice is $H_2O$ with lower fractions of other volatile molecules such as CO, $CO_2$, $NH_3$, $CH_4$, and $CH_3OH$ [for reviews see (Allamandola, et al. 1999; Burke & Brown 2010; van Dishoeck 2014)]. The first stage of molecular ice formation on cold grain surfaces is dominated by $H_2O$ ice and mixtures with $H_2O$ in all Galactic and extragalactic environments (Boogert, Gerakines, & Whittet 2015). Water ice plays an important role in the chemistry of the ISM, acting as a catalyst for chemical reactions and for the sticking of small dust particles. Concerning the role of water in space, we refer the reader to the following reviews (Bergin & van Dishoeck 2012; van Dishoeck et al. 2014).

Desorption of molecular ices increase the molecular diversity of the gas phase in different astrophysical environments including dense molecular clouds, protoplanetary and debris disks, envelopes of evolved stars, and surfaces and atmospheres of planets. It has been shown that for multicomponent ices based on water ice matrices, the desorption of all species in the ice is controlled by the behaviour of water (Collings et al. 2004; Martin-Doménech et al. 2014), thus, the desorption of water ice triggers the release of other molecules, which play an active role in the gas phase chemistry. In addition, the desorption properties of water ice are directly related to the observed amounts of gaseous and solid $H_2O$ in protoplanetary disks (Dominik et al. 2005; Min et al. 2016).

Thus, the study of the desorption of $H_2O$ ice mixed with dust particles is critical for our understanding of ice-dust interactions, the structure and morphology of dust grains and comets, the chemical diversity of the gas phase and thereby the physics and chemistry of different cosmic environments. A number of laboratory experiments studying the desorption of ices *from* dust surfaces have been performed. Here, we refer the reader to a review paper (Burke & Brown 2010) and recent studies (Noble et al. 2012; Dawley, Pirim, & Orlando 2014; Shi, Grieves, & Orlando 2015; Suhasaria, Thrower, & Zacharias 2017). However, as it was discussed above, dust in dense molecular clouds, planet-forming disks, and comets is mixed with molecular ices. Therefore, real astronomical dust aggregates are better modelled by ice *mixed with* dust rather than ice *layered* on dust. To our knowledge, the desorption of ices *mixed with* dust grains has not yet been reported. In the present study, we aim to provide results on the temperature-programmed desorption (TPD) of water ice mixed with laboratory analogues of interstellar and circumstellar amorphous carbon or silicate particles. The results are relevant for planet-forming disks, where a central star can trigger the thermal desorption of molecular ices.

## 2. Experimental part

Dust/ice mixtures were produced in a setup combining a laser ablation experiment with a cryogenic chamber. In this set-up, solid carbonaceous and siliceous grains can be condensed together with relevant gas molecules at temperatures down to 8 K. The setup (without the cryostat) is described in detail elsewhere (Jäger et al. 2008).

In our experiments simulating the formation of cosmic dust, the deposition of nanometre-sized hydrogenated fullerene-like amorphous carbon and amorphous silicate ($MgSiO_3$ and $MgFeSiO_4$) particles was performed by pulsed laser ablation of graphite, MgSi, and MgFeSi targets and subsequent condensation of the evaporated species in a quenching atmosphere of either $He/H_2$ (volume ratio of 5/2) for carbon grains or $O_2$ for silicate grains. The pressure in the ablation chamber was kept at 4 mbar. The low pressure regime applied is comparable to the pressure conditions for dust condensation in AGB stars (Lodders & Fegley 1999). The condensed dust grains were extracted adiabatically from the ablation chamber through a nozzle into a second chamber which was held at a pressure of around $10^{-3}$ mbar. During such an expansion, the particles decouple from the gas. Consequently, further particle growth due to reactions with reactive gas molecules and agglomeration processes are prevented. A second extraction was performed through a skimmer into a third chamber with a pressure of $10^{-6}$ mbar. The generated particle beam was directed into a fourth, cryogenic chamber, where the dust grains were deposited onto a KBr substrate cooled down to 8 K. Water ice molecules evaporated from a water reservoir and introduced through a leakage valve and a capillary tube were simultaneously deposited with grains. The deposition rate was varied from 2 to 10 nm/min in order to adjust the ice and grain deposition. Ice deposited at such low rates and low temperatures, forms high-density amorphous solid water (ASW) (Hagen, Tielens, & Greenberg 1981; Berland et al. 1995; Jenniskens et al. 1995). All depositions were performed in a vacuum chamber with a base pressure of $10^{-7}$ mbar. Such relatively poor vacuum conditions due to the combination of a laser ablation system with the cryogenic chamber lead to a deposition of $CO_2$ from the chamber atmosphere ($CO_2$ stretching band is visible in the IR spectra recorded). However, the amount of $CO_2$ is small compared to the amount of $H_2O$ and cannot influence the kinetics of desorption of water ice discussed in this study.

The thickness of the grain deposits was controlled by a quartz crystal resonator microbalance (sensitivity 0.1 nm) using known values for the deposit area and density. The thickness of the water ice was calculated for all depositions from the 3.07 μm water band area using a band strength of $2\times10^{-16}$ cm/molecule (Hudgins et al. 1993). The carbon/ice mass ratio was varied between 0.1 and 1.3. The silicate/ice mass ratios were 9.4 for the $MgSiO_3$ and 13.8 for the $MgFeSiO_4$ samples. The mass ratios were calculated from the thicknesses of the deposits using densities of 1.1 g cm$^{-3}$ for high-density amorphous water ice (Narten 1976), 1.7 g cm$^{-3}$ for amorphous carbon (Sabri et al. 2015), 2.5 g cm$^{-3}$ for amorphous $MgSiO_3$ silicates (slightly lower value as 2.7 g cm$^{-3}$ measured for glassy silicates (http://www.astro.uni-jena.de/Laboratory/OCDB/index.html)), and 3.7 g cm$^{-3}$ for amorphous $MgFeSiO_4$ silicates (Jäger et al. 2016). The thicknesses and mass ratios used are presented in Table 1.

TPD experiments were performed by linear ramping of the substrate temperature with a rate of 1 K/min in the temperature range between 8 and 200 K. The error of the temperature measurements was determined to be ±1 K. Infrared spectra during the warming-up were measured using a Fourier transform infrared (FTIR) spectrometer (Vertex 80v, Bruker) in the spectral range from 6000 to 400 cm$^{-1}$ with a resolution of 1 cm$^{-1}$. 16 scans were taken for one spectrum, which was equivalent to 40 seconds of the scanning time.

Table 1. Dust and ice thicknesses and dust/ice mass ratios.

| Sample number, N | Sample description | Dust/Ice mass ratio |
|---|---|---|
| 1 | Pure H$_2$O ice, 450 nm | |
| 2 | Pure H$_2$O ice, 40 nm | |
| 3 | Carbon grains, 35 nm + H$_2$O ice, 670 nm | 0.1 |
| 4 | Carbon grains, 35 nm + H$_2$O ice, 320 nm | 0.2 |
| 5 | Carbon grains, 35 nm + H$_2$O ice, 130 nm | 0.5 |
| 6 | Carbon grains, 30 nm + H$_2$O ice, 40 nm | 1.3 |
| 7 | MgSiO$_3$, 150 nm + H$_2$O ice, 40 nm | 9.4 |
| 8 | MgFeSiO$_4$, 75 nm + H$_2$O ice, 20 nm | 13.8 |

## 3. Results

The obtained vibrational spectra of H$_2$O ice are dominated by a broad band around 3300 cm$^{-1}$, which involves symmetric and antisymmetric OH stretching vibrations. IR spectra of a H$_2$O ice and a carbon/ice mixture are presented in Figure 1. One can see five water bands related to the stretching and bending vibrations at 3300 cm$^{-1}$ and 1660 cm$^{-1}$, to the librational motion at 750 cm$^{-1}$, to a combination mode at 2200 cm$^{-1}$, and to the presence of free H$_2$O molecules at 3680 cm$^{-1}$. In addition, the stretching band of CO$_2$ at 2344.9 cm$^{-1}$ is visible due to the relatively poor vacuum conditions used.

A number of differences between the spectrum of the carbon/ice mixture and the spectrum of pure water ice presented in Figure 1 can be observed. The spectral properties of dust/ice mixtures at different temperatures are outside of the scope of the present paper. The optical properties of silicate/water ice mixtures have been published recently (Potapov, et al. 2018) and a study of the optical properties of carbon/water ice mixtures is in progress.

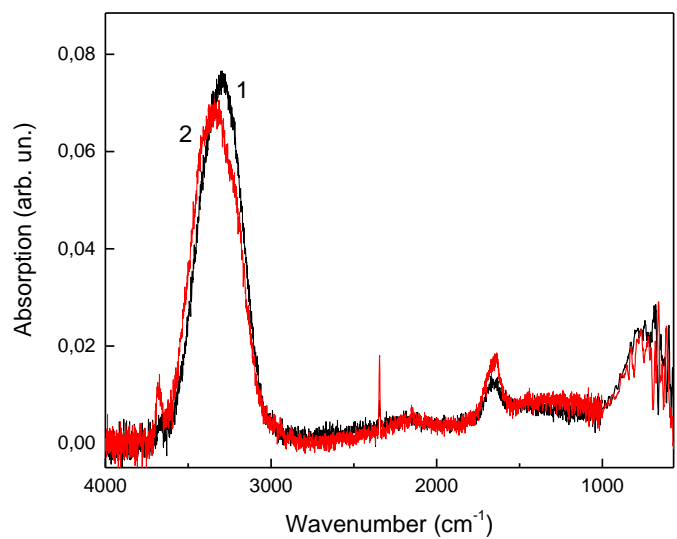

Figure 1. IR spectra at 8 K: 1 - H$_2$O ice (40 nm, N2 in Tables 1 and 2) and 2 - carbon grains/H$_2$O ice (30 nm/40 nm, N6 in Tables 1 and 2). The spectra were smoothed in the range below 1000 cm$^{-1}$.

The high-density ASW converts into the low-density ASW in the temperature range between 30 and 80 K and then into crystalline ice (CI), typically, in the temperature range between 130 and 150 K (Hagen, et al. 1981; Jenniskens, et al. 1995; Fraser et al. 2001; Bolina, Wolff, & Brown 2005a). Both transformations lead to changes in the water band positions, shapes, and intensities (Hagen, et al. 1981; Jenniskens, et al. 1995). Following the evolution of the 3300 cm$^{-1}$ water ice band with the temperature, we observed the spectral changes corresponding to both structural transformations, specifically, a redshift, a narrowing, an intensity increase and the appearance of a substructure of the band.

TPD curves were obtained by heating the samples with a heating rate of 1 K/min and taking the first temperature derivative of the integrated intensity of the OH stretching band. In Figure 2, TPD curves for pure water ices of two different thicknesses are shown. Figure 3 presents a TPD curve for pure H₂O ice together with a TPD curve for the carbon grains/water ice sample with the same amount of water. The TPD curves of all carbon/ice samples are shown in Figure 4.

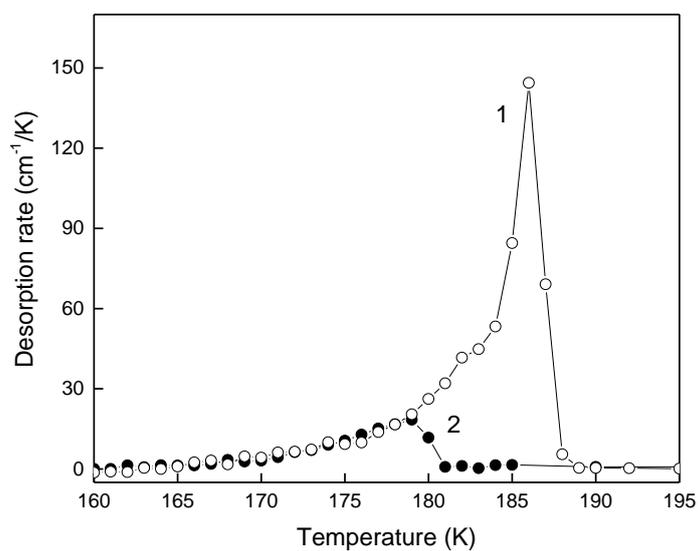

Figure 2. TPD curves for $H_2O$ ice: 1 - 450 nm (N1 in Tables 1 and 2) and 2 - 40 nm (N2 in Tables 1 and 2).

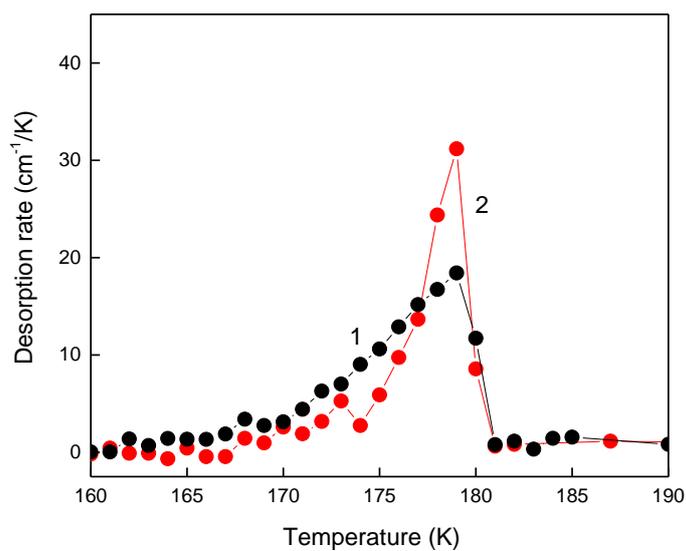

Figure 3. TPD curves: 1 - $H_2O$ ice (40 nm, N2 in Tables 1 and 2) and 2 – carbon/ice (30 nm/40 nm, N6 in Tables 1 and 2).

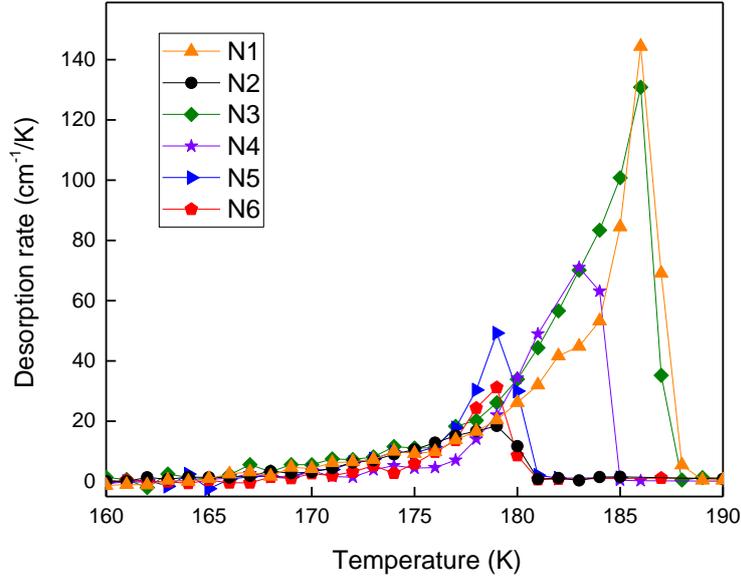

Figure 4. TPD curves for the samples N1 − N6 (see Tables 1 and 2).

Following a number of studies (Acharyya et al. 2007; Noble, et al. 2012; Martin-Doménech, et al. 2014), we fitted the leading edges of the TPD curves with the Polanyi-Wigner equation (Polanyi & Wigner 1928):

$$\frac{dA}{dT} = v_0 \frac{1}{b} N^i e^{\frac{-E}{RT}} \qquad (1)$$

where *A* corresponds to the integrated band intensity, *T* is the temperature, $v_0$ is the pre-exponential factor of $10^{12}$ s$^{-1}$, which is a standard value for physisorbed species (Sandford & Allamandola 1988; Biham et al. 2001; Fayolle et al. 2011), and *b* is the heating rate of 1 K/min. *N* and *i* correspond to the number of molecules and to the desorption order, respectively. *E* represents the desorption energy and *R* the gas constant. The pre-exponential factor depends on the ice thickness, but this dependence is not strong, even for non-zero desorption orders (Bisschop et al. 2006; Martin-Doménech, et al. 2014). As it was shown, an effect of altering the desorption energy on the pre-exponential factor for water ice is very small (Bolina, et al., 2005a). Therefore, the pre-exponential factor was kept constant for all the studied samples. The results of the fits of the TPD curves presented in Figure 3 are shown in Figure 5 a) and b).

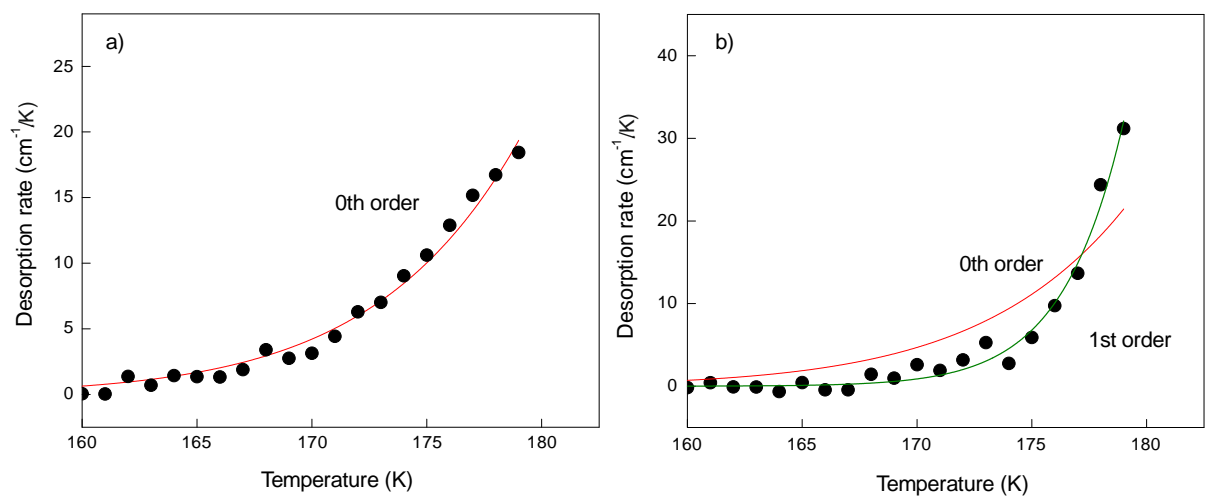

Figure 5. Leading edges of the TPD curves: a) pure $H_2O$ ice (N2 in Tables 1 and 2) and b) carbon/ice sample (N6 in Tables 1 and 2) and their fits using $0^{th}$ and $1^{st}$ desorption orders.

For pure water ice multilayer (40 nm of water ice is more than 100 monolayers) we obtained $0^{th}$ order of desorption in agreement with the literature (Fraser, et al. 2001; Bolina, et al. 2005a). In addition, an alternative approach for the determination of the desorption order proposed in (Bolina, et al. 2005a), from the gradient of the plot of the natural logarithm of the TPD peak versus the relative ice coverage at a fixed temperature, was used. The zeroth desorption order of the pure water ice was confirmed.

For the carbon/ice sample N6, the result is different. In order to fit the leading edge of the TPD curve, we had to use $1^{st}$ order desorption kinetics (see Figure 5b). A series of experiments using different carbon/ice mass ratios was performed. Variations of these ratios lead to a transformation of the TPD curve, which can be fitted with the Polanyi-Wigner equation using fractional desorption orders. Such an approach was used to describe TPD curves for $H_2O$, $NH_3$, and $CH_3OH$ ices (Bolina, et al. 2005a; Bolina, Wolff, & Brown 2005b; Brown & Bolina 2007). The desorption order increases with an increase of the carbon/ice mass ratio as shown in Table 2.

Table 2. Dust/ice mass ratios, desorption orders used to fit the TPD curves,

peak desorption temperatures, and desorption energies.

| N | Dust/Ice mass ratio | Desorption order | Desorption temperature (K) | Desorption energy (kJ/mol) |
|---|---|---|---|---|
| 1 | – | 0 | 186 | 48.7 ± 0.3 |
| 2 | – | 0 | 179 | 48.7 ± 0.3 |
| 3 | 0.1* | 0.3 | 186 | |
| 4 | 0.2* | 0.7 | 183 | |
| 5 | 0.5* | 1.0 | 179 | 49.7 ± 0.5 |
| 6 | 1.3* | 1.0 | 179 | 49.7 ± 0.5 |
| 7 | 9.4** | 1.0 | 178, 183 | 49.4 ± 0.5, 50.9 ± 0.5 |
| 8 | 13.8** | 1.0 | 177, 180 | 49.1 ± 0.5, 50.0 ± 0.5 |

Notes. – Pure water ice, *carbon/ice mixtures, **silicate/ice mixtures

The desorption energy obtained for the pure water ice using equation 1 is 48.7 ± 0.3 kJ/mol in agreement with literature data (Fraser, et al. 2001; Collings, et al. 2004). Redhead's peak maximum method (Redhead 1962) valid for the first order desorption gave us a value for the desorption energy $E$ of 49.7 ± 0.5 kJ/mol for the carbon/ice samples N5 and N6. All desorption energies determined are presented in Table 2.

In addition, we present our first experimental data on silicate/ice mixtures. TPD curves of the $MgSiO_3$/$H_2O$ (N7) and $MgFeSiO_4$/$H_2O$ (N8) samples are shown in Figure 6.

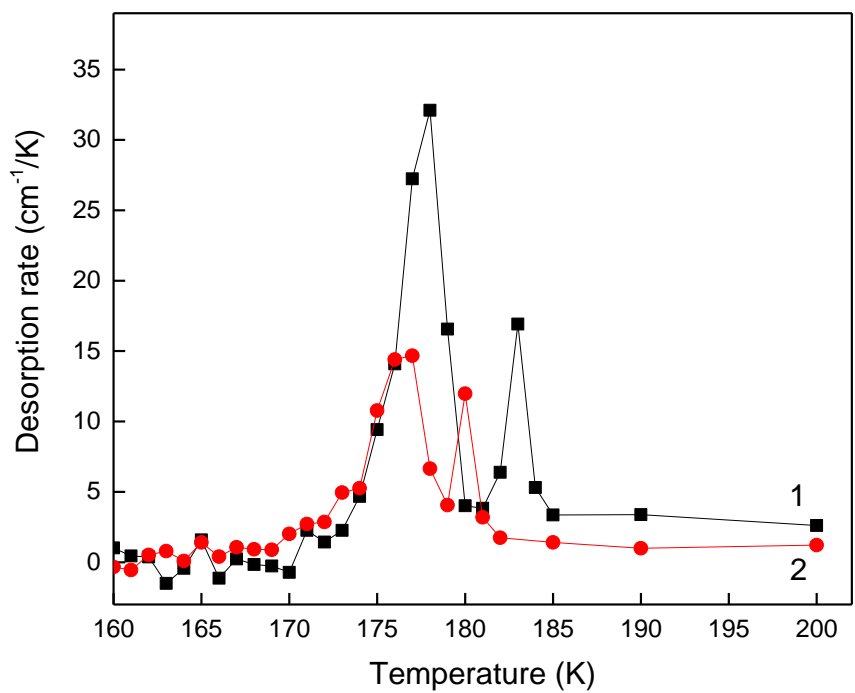

Figure 6. TPD curves: 1 - MgSiO$_3$/H$_2$O sample (N7 in Tables 1 and 2) and 2 - MgFeSiO$_4$/H$_2$O sample (N8 in Tables 1 and 2)

As one can see, there is a principal difference between the silicate/ice and carbon/ice TPD curves, namely the presence of a second peak in the silicate/ice measurements. In addition, we observed slightly increased desorption rates beyond the second desorption maxima indicating a rest of water ice in the samples up to 200 K. This result is in agreement with our recent study, which showed a trapping of water ice molecules in silicate grains (Potapov, et al. 2018). For both silicate/ice samples, first order kinetics of desorption was determined (Figure 7 a, b). The desorption energies obtained using the Readhead's method are presented in Table 2.

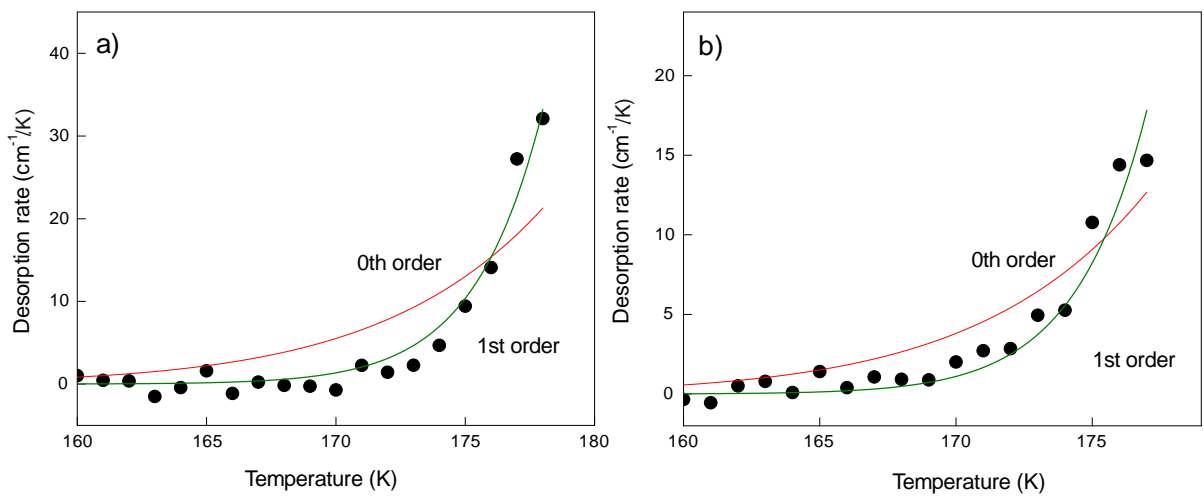

Figure 7. Leading edges of the TPD curves: a) MgSiO$_3$/H$_2$O sample (N7 in Tables 1 and 2) and b) MgFeSiO$_4$/H$_2$O sample (N8 in Tables 1 and 2) and their fits using 0$^{th}$ and 1$^{st}$ desorption orders.

## 4. Discussion

<u>4a. Kinetics of desorption. Desorption of water ice molecules from the surface of fractal clusters of dust grains.</u>

Why water ice mixed with dust grains shows different kinetics of desorption in comparison with the same amount of pure ice? This result has not been obtained in previous experiments on the thermal desorption of ices *from* carbon and silicate surfaces. In the following we will try to answer the question stated above.

Fractional desorption orders in molecular ices, such as H$_2$O, NH$_3$, and CH$_3$OH, has been attributed to the hydrogen-bonded network, which results in a strong interaction between adjacent molecules. Therefore, the desorption of one molecule is not independent of the desorption of neighbouring molecules, as would be expected for zero-order desorption (Bolina, et al. 2005a,2005b; Brown & Bolina 2007). Similarly, the formation of molecular clusters in glycolaldehyde and acetic acid ices has been shown to influence their kinetics of desorption (Burke et al. 2015) due to the intermolecular hydrogen bonds formed between the ice molecules, even at the lowest surface coverage. However, the desorption orders of all of these ices in the multilayer regime were determined to be close to zero. In our case, the desorption order rises up to 1, so hydrogen-bonded network between water molecules is, probably, not an appropriate answer.

Another answer could be an influence of the structure of water ice deposited with or without grains on the desorption behaviour. It was observed that a part of the desorption of water ice after the ASW – CI phase transition occurs from the residual amorphous phase showing a shoulder in the leading edge of a TPD curve (Fraser, et al. 2001). The authors showed that the weak TPD peak related to the desorption of amorphous solid water is at lower temperature than the main TPD peak related to the desorption of crystalline ice, due to the higher vapour pressure of the amorphous ice, and does not noticeably alter the curve behaviour. In the same study, a comparison between TPD dependencies of high and low density ASW was performed and the curves were found to be essentially equivalent. Thus, the structure of ice is also not an answer that we are looking for.

Typically, zeroth-order desorption relates to a desorption of ice multilayers and the desorption rate does not depend on the number of adsorbed molecules. First-order desorption arises from a monolayer and sub-monolayer desorption and the desorption rate depends linearly on the number of molecules on the surface. Second-order desorption refers to a desorption of chemically formed species or to a specific distribution of binding sites on the surface when a desorption of one molecule causes a desorption of another one. In this case, the desorption rate depends quadratically on the number of adsorbed molecules. The first-order desorption of water ice in the dust/ice mixtures could be explained by the large surface area of dust clusters that form during the dust deposition. $H_2O$ molecules occupy this surface forming monolayer or sub-monolayer rather than thick water ice multilayer as in the case of pure water ice covering a flat substrate. The monolayer desorbs directly from the surface of dust grains following the first order kinetics. When the concentration of the dust in the ice is decreased, the ice coverage exceeds the dust surface and forms, probably, a monolayer with multilayer islands. Such an island structure can explain the fractal desorption orders related to the desorption of a monolayer and multilayers of ice in this case. The desorption order decreases with the increasing area of multilayer islands and approaches the desorption order of pure $H_2O$ ice. A link between a fractional order desorption kinetics and an island growth on the surface was proposed also elsewhere (Suhasaria, et al. 2017).

We have measured the structural characteristics of the deposited hydrogenated carbon grains by using high-resolution transmission electron microscopy (HRTEM) (Jäger, et al. 2008; Jäger et al. 2009). As an example, Figure 8 shows an HRTEM image of grains produced by laser ablation in 3.3 mbar He/$H_2$ quenching gas atmospheres.

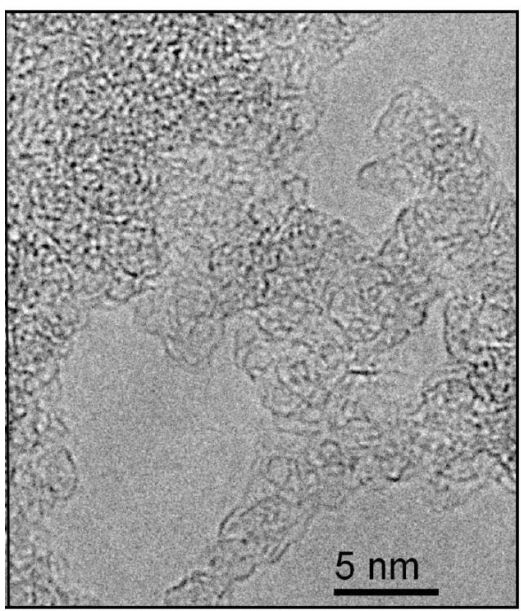

Figure 8. HRTEM image of carbon grains produced by laser ablation in 3.3 mbar He/$H_2$ quenching gas atmospheres (Jäger, et al. 2008).

The particles are composed of small, strongly bent graphene layers with varying lengths and distances between these layers. The individual particles are very small (less than 3-4 nm) and the largest particle agglomerates are in a range of 15 nm. The morphology of the deposit can be understood as a porous layer of rather fractal agglomerates. The measurements showed that the surface of the dust is very large, but the area cannot be measured exactly. This surface defines the thickness of the coverage of dust grains by water ice molecules.

Numerical simulations of the structure of cometary grain aggregates showed that they are fractals with low densities and irregular shapes [see (Fulle & Blum 2017) and references therein]. The fractal dimension of about 2 for small cometary grain aggregates was proposed in a number of studies (Donn & Hughes 1986; Samson, Mulholland, & Gentry 1987; Meakin & Donn 1988; Meakin, Donn, & Mulholland 1989; Donn 1990; Weidenschilling 1997; Kataoka et al. 2013; Tazaki et al. 2016) and as a result of collisional experiments (Wurm & Blum 1998; Krause & Blum 2004). Fractal dimension of 2 in 3D space means that physical and chemical processes occur on the *surface* of fractal aggregates supporting our suggestion about the desorption of ice from the surface of dust grains.

4b. Desorption temperature and energy. Trapping of water ice molecules in silicates

As one can see from Table 2, the desorption energies for dust/ice mixtures are only slightly higher than the one for pure water ice. The peak desorption temperature depends on the amount of ice deposited. It is the same for the samples N2, N6, and N7 (first peak) containing the same amount of water (see Table 1). The results indicate that ice molecules are equivalently bound to neighboring ice molecules and to the carbon surface. This is also true for the major part of water molecules interacting with the silicates. However, in the TPD curves of the silicate/ice samples, additional peaks at 183 K and 180 K point to differently bound ice molecules to the silicate surface. The presence of two desorption maxima was registered for CO, $N_2$, $O_2$, and $CH_4$ ices (Collings, et al. 2004; Suhasaria, et al. 2017). This finding was explained by the multilayer and monolayer desorption of ice, which can be discussed with stronger bonds between the species and the surface of the substrate compared to the bonds in the ice matrix. In contrast, $H_2O$, $NH_3$, $CH_3OH$, and $HCOOH$ ices were found to show only a single TPD peak (Collings, et al. 2004). According to the discussion in the previous section, a multilayer desorption of water ice from the silicate samples can be excluded. Thus, in our study, the two peaks do not correspond to the multilayer and monolayer desorption as in the case of CO, $N_2$, $O_2$, and $CH_4$ ices, but to water molecules differently bound to silicate grains.

The residual desorption beyond the TPD maxima shows that a small part of ice molecules might be physically trapped in small pores of silicate aggregates even at temperatures beyond the desorption temperature of pure water ice. In addition, it can correspond to the release of chemically trapped OH groups from the surface of the silicates. The trapping of water ice at temperatures up to 200 K was also observed in our recent study (Potapov, et al. 2018). The trapping of ice molecules was not detected for carbon/ice mixtures, although carbon grains are also very porous.

The difference between carbon and silicate grains can point to the hydrophobic and hydrophilic properties of their surfaces. The transition from hydrophobic to hydrophilic behavior of carbon surfaces due to the formation of oxygenated sites has been simulated (Müller & Gubbins 1998). Hydrophobic-to-hydrophilic conversion by oxidation applied to carbonaceous aerosol (Petters et al. 2006) and carbon nanowalls (Watanabe et al. 2012) is discussed in the literature. The fullerene-like carbon grains used in the present study are characterized by strongly bent graphene layers linked by aliphatic chains and therefore they are typically considered as a hydrophobic material. In our experiments, oxygenation of the carbon surface did not occur. On the other hand, silica ($SiO_2$) as well as silicates are known to have hydrophilic and hydrophobic surface groups (Klier, Shen, & Zettlemo 1973; Lee & Rossky 1994; Yu et al. 2018). The structure of silicates is characterized by $SiO_4$ tetrahedra, which can be isolated or linked with adjacent tetrahedra by oxygen bridges. The polymerization degree of the tetrahedra depends on the stoichiometry of the material. At the surface of silicates, either hydrophilic silanol groups (Si-OH) or hydrophobic Si-O-Si groups are present. The ratio of these groups depends on the production process, the thermal treatment of the silicates, and on a possible processing of silicate surfaces. In silicates produced by laser ablation, the amount of Si-OH groups is small. However, reactions between water molecules and grains have probably increased the number of silanol groups on the surface of the grains. Consequently, the number of hydrophilic surface sites was increased.

Thus, with our results, we demonstrate a principal difference between the surface properties of two classes of interstellar grains, namely, the finding of differently bound water ice molecules to silicate grains in contrast to fullerene-like carbon grains. However, this is not a general statement for carbon because carbon surfaces can be very different. For example, oxygenation of C-atoms can produce hydrophilic surface sites.

4c. Astrophysical implication

The knowledge of the desorption properties of water ice mixed with dust is very important for the understanding of astronomical observations of water vapour in planet-forming regions and for the revealing of the architecture and evolution of planet-forming disks. The desorption of water ice triggering the release of other volatiles trapped in the ice influences the chemistry of the gas phase of disks by feeding it with molecules. In addition, the formation efficiency of larger bodies depends on the ice coverage of dust grains.

Knowing the amount of ice relative to dust in a planet-forming disk and the kinetics of desorption of the ice, the structure and morphology of grains and larger bodies can be modelled. It will help to trace back the thermal history of the disk. Conversely, knowing the thermal history of a cosmic body seeded with molecular ice and the kinetics of desorption of ice, the abundance of molecules stayed in the ice and released into the gas phase can be estimated. It will improve our understanding of astrochemical processes in the gas phase and on surfaces. However, more systematic studies on the thermal desorption of molecular ices mixed with different dust analogues are needed. This work represents a first small step in this challenging direction.

**Summary**

We measured the temperature-programmed desorption of water ice, which had been premixed with amorphous carbon or silicate dust grains, and demonstrate the dependence of the kinetics of desorption of water ice on the dust/ice ratio. The desorption order increases with an increase of the dust/ice mass ratio and reaches 1, indicating the desorption of ice molecules from the large surface of dust grains. This result shows an important link between the structure and morphology of icy dust grains in planet-forming disks and the amount and the desorption properties of seeded ice. If more information on the thermal history of dust grains was known, more information about the structure and morphology of these grains in the interstellar medium and planet-forming disks could be derived. Moreover, a principal difference between the surface properties of two classes of interstellar grains, namely, the finding of differently bound water ice molecules to silicate grains in contrast to fullerene-like carbon grains is demonstrated.

**Acknowledgments**

This study has greatly benefitted from discussions with Guillermo Muñoz Caro and Patrice Theulé. It was supported by the Research Unit FOR 2285 "Debris Disks in Planetary Systems" of the Deutsche Forschungsgemeinschaft (grant JA 2107/3-1).